# When physics meets chemistry at dynamic glass transition

Haibao Lu

Science and Technology on Advanced Composites in Special Environments Laboratory, Harbin Institute of Technology, Harbin 150080, China

[*]Corresponding author, E-mail: luhb@hit.edu.cn

## (i) an explanation of why the topic is of general interest

Can the laws of physics be unified? One of the most puzzling challenges is to reconcile physics and chemistry, where molecular physics meets condensed-matter physics, resulting from the scaling effect and dynamic fluctuation of glassy matter at the glass transition temperature. Pioneer of condensed-matter physics, the Nobel Prize-winning physicist Philip Warren Anderson, wrote in 1995: The deepest and most interesting unsolved problem in condensed-matter physics is probably the theory of the nature of glassy state and the glass transition. In 2005, the question of "what is the nature of glassy state?" was suggested as one of the greatest scientific conundrums over the next quarter-century for Science's 125th anniversary. However, the nature of glassy state and its connection to the glass transition have not been fully understood owing to the interdisciplinary complexity of physics and chemistry, where they are governed by the physical laws at condensed-matter and molecular scales, respectively. Therefore, study on the glass transition becomes essential to explore the working principles of scaling effect and dynamic fluctuation in glassy matter, as well as further reconcile the interdisciplinary complexity of physics and chemistry.

(1) The glass transition of glassy matter has been identified to follow the dynamic

fluctuation of chaos, where the molecular dynamics of glassy matter is derived from the long-term structure and ruled by the condensed-matter cohesion, of which the intermolecular force is too small to achieve it amorphous. Based on the transition probability of Bernoulli distribution for arrangement and combination, the glass transition has been identified as a dynamic transition from order to disorder, of which the equilibrium relationship has been formulated using the order and disorder free-energies of thermodynamic microphase separation. It has been revealed that the dynamic equilibrium of the glass transition originates from the interplay of phase and microphase separations at condensed and molecular scales, owing to inter- and intra-molecular cohesions, respectively, resulting into dynamic fluctuation, scaling effect and complexity. Furthermore, the Adam-Gibbs (AG) domain model has been employed to explore the cooperative dynamics and molecule entanglement in glassy matter, based on the transition probability of pairing problem, where there are constant $e+1 \approx 3.718$ segments in a domain to relax cooperatively at the glass transition temperature. Finally, the claim of "50~100 monomers relaxing synchronously at glass transition temperature", for the first time, has been theoretically modeled and verified.

(2) The connection of glass transition to free volume has been identified as a constitutive relationship between dynamic chaos and mathematic fractal for the glassy matter. In this relationship, the free volume works as an aid to the dimensional fractal for glass transition, which is then considered as a temporal chaos for free volume. Instead of a Flory-Fox's phenomenology assumption, physical insight toward the free volume is derived from Einstein's mass-energy equation. Meanwhile, the mathematic insight into free volume originates from fractal theory, which uses iterative equations to predict the trail of free volume following the self-consistency and power law. However, differences in fractal dimensions between the molecular chain and their condensed aggregate have been identified as the driving force behind the scaling effect and dynamic fluctuation. Finally, the dynamic equilibrium of free volume has been formulated using the thermodynamic separation of dense and dilute phases. These phases are used for the glassy matter and glassy matter-free volume parts, respectively, where the free volume works as the solvent. The constant free-volume fraction of 2.48% is phenomenologically obtained to achieve the condensed constant of $0.12(1-\gamma)$ (where $\gamma$ is the free-volume's overlap constant) at the condensed-matter scale, analogous to the Boltzmann constant ($k_B$) and gas constant ($R$).

(3) This study is expected to provide a fundamental framework to explore the nature of the glassy state and formulate the dynamic equilibrium of glass transition. The dynamic fluctuation and scaling effect are originated from the interplay of interdisciplinary complexity at condensed matter and molecular scales, where physics meets chemistry, resulting into their reconciliations achieved to unify the laws of physics at condensed-matter scale.

### (ii) A brief outline of the article's contents

### 1. Introduction

Figure 1 shows, in a unified way, a selection of length scale of all physics, i.e., electron at $10^{-18}$ m, atomic nucleus at $10^{-15}$ m, atom at $10^{-10}$ m, molecule at $10^{-9}$ m, macromolecule at $10^{-7}$ m, condensed-matter at $10^{-4}$ m, classical mechanics at $10^{2}$ m,

Geophysics at $10^8$ m and astrophysics at $10^{26}$ m [1]. Note that length scales in physical system have become an essential skill to create predictive simulations across multiple length scales, and to further answer the question of "can the laws of physics be unified?" [1-6]. Therefore, it is important to get an idea of how big differences in different length scales [2,3,5]. An example is the Bohr model of single-electron atoms, in which the electron is pictured as orbiting the nucleus in a similar way to how planets orbit the Sun [7]. The difference might be the scaling effect between atomic physics and Geophysics [8]. Furthermore, the scaling effect of condensed matter is expected to originate from the interplay of inter-chain phase and intra-chain microphase separations.

[Figure 1]

Here, the scaling effects of two distinct length scales are essential to reconcile the interdisciplinary laws, as shown in Figure 2. However, studies on scaling effects are challenging to promote, as they originate from complex and interdisciplinary systematologies [9,10]. Dynamic fluctuations can suddenly occur [2], resulting in a non-equilibrium and discontinuous but self-consistent scaling effect. Note that there is a scaling effect for glassy matter undergoing dynamic glass transition [3,5], at two condensed and molecular length scales. Since molecular physics has been recognized as a part of chemistry, it is possible to achieve the scaling effect of dynamic glass transition in glassy matter based on the reconciliation of molecular physics (i.e., chemistry) and condensed-matter physics.

[Figure 2]

The nature of the glassy state is probably the most intricate issue in condensed matter physics [11]. Even for crystalline matters, it is difficult to achieve a 100% degree of crystallinity, resulting in the existences of amorphous matters in them. In 2005, the question "What is the nature of glassy state?" was suggested as one of the greatest scientific conundrums over the next quarter-century for Science's 125th anniversary. Studies on the nature of glassy matter have been ongoing for several decades, due to both the fundamental significances and broad implications in condensed matter physics [12-16]. The motivations for studying glassy matters are numerous, and the reasons why this research field having received increasing attention have been explained through numerous enumerations [17]. In condensed matter physics, the nature of glassy matter and its connection to the glass transition is one of the most essential and fascinating topics. This is because almost all condensed matters originates from its glassy states, resulting in an inseparable connection between dynamics and the glass transition. However, theoretical efforts remain at the phenomenological level, according to the empirical results, and the actual scientific explanation behind it has yet to be revealed [18].

In this study, a thermodynamic order-to-disorder free-energy equation for microphase separation has been proposed based on the transition probability. This equation formulates the dynamic equilibria for the glass transition, where the fluctuation originates from the interplay of phase and microphase separations at two scales: condensed and molecular. It is used to achieve a reconciliation of physics and chemistry at both the condensed and molecular scales. Moreover, the Adam-Gibbs

(AG) model has been introduced to explore the dynamic cooperativity of a constant domain size $z=e+1$ at the glass transition based on the principle of maximum entropy (principle of minimum entropy production) and transition probability of pairing issue, and verify the experienced claim of "50~100 monomers relaxing synchronously at glass transition". Additionally, the connection between the glass transition and free volume, which originates from Einstein's mass-energy equation, has been identified as a constitutive relationship between dynamic chaos and fractals. Then the complex fractal function has been introduced to predict the trail guild of free volume. This function follows self-consistency and power laws, where the scaling effect is a result of the differences in fractal dimensions between the molecule chain and condensed matter. A free-volume equilibrium has been developed using the dynamic separation of dense and dilute phases. This equilibrium verifies that the glass transition occurs at a constant fraction of free volume at 2.48%, which is then used to achieve the condensed constant of $C = 0.12 \times (1-\gamma)$ (where $\gamma$ is the free-volume's overlap constant) at the condensed-matter scale, analogous to the Boltzmann and gas constants. Finally, this study provides a fundamental framework for understanding the nature of glassy matter and formulating the dynamic equilibrium of the glass transition. The dynamic fluctuation originates from the interplay of interdisciplinary complexity at two scales: condensed and molecular, where physics meets chemistry.

## 2. Connection of glass transition to the transition probability

### 2.1 Dynamic equilibrium of glass transition

The Eyring equation has been previously combined with the Williams-Landel-Ferry

(WLF) equation to characterize the thermodynamics of the glass transition in amorphous polymer [19]. In the glassy state, the relaxations of molecular chains require sufficient activation energies to overcome their energy barriers. On the other hand, in the rubbery state, molecular chains require free volume to be available in order to jump, as no further energy barrier exists to constrain their dynamic relaxations [20]. Therefore, the transition probability of a molecular chain to relax ($p_j$) around the $T_g$ can be simply written as:

[Equation (1)]

where $p_e(T)$ is the transition probability of a molecular chain to obtain sufficient activation energy to overcome the energy barrier, and $p_v(T)$ is the transition probability to obtain enough free volume available to jump.

Based on the Binomial distribution (Bernoulli distribution) with parameters *n* and *p*, the discrete distribution in a sequence of *n* independences in statistics. Here one single step is called a Bernoulli trial, and a sequence of steps forms a Bernoulli process. Therefore, the transition probability of molecular chains to relax is governed by the mathematical arrangement of the Bernoulli process below $T_g$, where the random variable (*p(X)*) is 1 or 0 depending on if the molecular chain is able to relax or not. As shown in Figure 3(a), the constitutive relationship between distribution function ($F(M_i)$) and molecular weight ($M_i$) is plotted based on the transition probability of Bernoulli distribution for the glassy matter. Therefore, below the $T_g$, glassy matter presents a transition probability of molecular arrangement, and molecular chains do not relax unless enough activation energy is obtained [19].

[Figure 3]

Consequently, above the $T_g$, there is a transition probability of combination, where molecular chains have all obtained enough activation energies to relax. As is well known, physical order and disorder can be represented by arrangement and combination in mathematics, respectively [21]. Therefore, based on the probability of Bernoulli distribution function ($F(X)=\Sigma p(X \leq M_i)$), where $M_i$ is the molecular weight, the glass transition of glassy matter is a physical order-to-disorder or mathematic arrangement-to-combination issue, as shown in Figure 3(b).

As revealed in Figure 4, the relaxation behavior of glassy matter is derived from the molecular chain below the $T_g$, while it originates from the molecular segment above the $T_g$. Therefore, the glass transition has been identified as the temperature point at which segmental relaxation in a molecular chain begins [22,23]. The relaxation motion of a molecular chain has to overcome the intermolecular constraints of other chains in their condensed states and is governed by the thermodynamics of phase separation [24,25]. Meanwhile, the relaxation motion of a segment has to overcome the intramolecular constraints of other segments in a molecular chain and is governed by the thermodynamics of microphase separation [24,25].

[Figure 4]

Therefore, the dynamic equilibrium of glass transition is derived from the ordered ($F_{ORD}$) and disordered ($F_{DIS}$) free energies of the segments undergoing microphase separation in a molecular chain. According to the thermodynamics of microphase separation, the ordered ($F_{ORD}$) free energy is expressed as [24,25],

[Equation (2)]

where $\gamma_{ORD} = \dfrac{k_B T}{b^2}\sqrt{\dfrac{\chi}{6}}$ is the interfacial tension [25], $\Sigma \dfrac{L}{2} = Nb^3$ is the interfacial area [24], $\chi$ is the interaction parameter between two microphases [22], $L$ is the interfacial thickness, $b$ is the segment length, $R=8.314$ J/(mol K) is the gas constant, $N = \dfrac{V_s}{V_m}$ ($\dfrac{V_s}{V_m} = \dfrac{Nb^3}{b^3} = N$) is the number of monomers in a dynamic segmental unit, where $V_s$ and $V_m$ are the volumes of segment and monomer, respectively [25]. If the lowest ordered free-energy is achieved, i.e., $\dfrac{\partial F_{ORD}}{\partial L} = 0$, $L_{opt} = bN^{2/3}\chi^{1/6}$ and $L_{opt}$ is the periodically interfacial thickness [24,25], the ordered free-energy function can therefore be expressed as,

[Equation (3)]

Meanwhile, the disordered free-energy of segment is employed as [24,25],

[Equation (4)]

where $\phi_O$ and $\phi_D$ are the volume fractions of order and disorder microphases in the polymer chain, respectively. At $T_g$, it is assumed that $\phi_O = \phi_D = \dfrac{N_O}{N_O + N_D} = \dfrac{1}{2}$, resulting from the Flory-Huggins solution theory, i.e., the volume fractions of two phases are equal to each other at critical point [23-25].

In combination with equations (3) and (4), the thermodynamic equilibrium of microphase separation can be then obtained as,

[Equation (5a)]

[Equation (5b)]

Here $\chi$ is also used to characterize the interaction parameter between the order and disorder microphases. Owing to the strongly chemical bonding in the molecule chain, it is chosen as $\chi \approx 0.5$, resulting into $N \approx 21$. That is to say, the basic segment containing $N = 21$ monomers undergoes microphase separation at $T_g$. Furthermore, Figure 5 plots the constitutive relationship of interaction parameter ($\chi$) as a function of number of monomers in a dynamic segmental unit ($N$). It is revealed that the number of monomers in a dynamic segmental unit ($N$) is gradually increased with a decrease in the interaction parameter ($\chi$), at the same $\frac{N_O}{N_O + N_D} = \frac{1}{2}$.

[Figure 5]

## 2.2 Dynamic cooperativity and molecule entanglement at glass transition

Similar to quantum entanglement in quantum mechanics, molecule entanglement does exist and has been identified using the dynamic cooperativity for glassy matter undergoing a glass transition. This is because the relaxation of each segment cannot be characterized and has been completely covered by the dynamic cooperativity of all segments in the AG domain system. Therefore, the molecule entanglement originates from the dynamic cooperativity in condensed matter. Due to molecule entanglement, the configurational entropy of a segment is equal to that of the domain in mathematics, while the glass transition of glassy matter is derived from the AG domain in physics, rather than a segment within it.

Although the dynamic equilibrium of glass transition has been formulated based on the thermodynamic microphase separation and order-to-disorder free-energy functions. The dynamic cooperativity or molecule entanglement has not been formulated for the glass transition of glassy matter. The AG domain model provides an effective approach to studying the dynamic cooperativity of glassy matter [26]. The Cooperatively rearranging region (CRR) has been developed to explain the dynamic and cooperative relaxation of glassy matter [27,28], where the number of segments in a CRR domain has been defined as the domain size ($z$), and all these segments relax cooperatively and simultaneously to overcome their potential energy barriers, which restrict the relaxations and rearrangements. Meanwhile, all the segments relax synchronously in a domain, and each of them has the same dynamic characteristics.

For 1 mol segments being involved into cooperative relaxation, the configurational entropy ($S_c$) can be written as [29],

[Equation (6)]

where $N_z$ is the number of domains and $k_B = 1.38 \times 10^{-23}$ J/K is the Boltzmann constant. $\Omega$ is the configurational number of a dynamic segment in the domain, which relax cooperatively and synchronously [27-29].

The relationship between the domain size ($z$) and the number of domain ($N_z$) for 1 mol basic dynamic segments can be obtained [30],

[Equation (7)]

where $N_A = 6.02 \times 10^{23}$ is the Avogadro's number.

According to the AG model, the activation energy of a domain is determined by the domain size ($z$). With an increase in temperature, the domain size ($z$) decreases, while the configurational entropy increases accordingly [27-31]. By substituting equation (6) into (7), a constitutive relationship between the domain size ($z$) and configurational

entropy ($S_c$) can be expressed as [31]:

$$\text{[Equation (8)]}$$

Here, $s^*$ is defined as the maximum value of configurational entropy of 1 mol dynamic segments, of which the relaxation is not mutually restricted by other ones at the temperature of $T^*$, resulting into $z=1$ [32]. Therefore, $s^*$ can be written as:

$$\text{[Equation (9)]}$$

By substituting equation (9) into (8), the domain size ($z$) can be obtained as,

$$\text{[Equation (10)]}$$

Based on AG model, the configurational entropy ($S_c(T)$) can be expressed by the domain size in the temperature range from $T_0$ to $T^*$, where $T_0$ is defined as the temperature if all the conformers are involved into one domain and the configurational entropy is equal to zero, resulting into $z=\infty$ [30-34]. On the other hand, the $z=1$ at $T^*$, where the cooperative relaxation of conformers becomes independent. In the temperature range from $T_0$ to $T^*$, there are infinite values in that interval of domain size ($z$), where $z$ is ranged from $1 \leq z \leq +\infty$. To identify the cooperative relaxation using the configurational entropy ($S_c(T)$), the transition probability of pairing problem has been employed to characterize the domain size ($z$). That is to say, the transition probability should meet the requirement of at least one basically dynamic segment starts to relax at the $T_g$, where the segments have to be paired with their corresponding molecule chains.

If there are $n$ ($0<n\leq+\infty$) molecule chains, correspondingly, there are $n$ basically dynamic segments to relax in these chains at $T_g$. Here there are two requirements for the transition probability at $T_g$, i.e., (1) at least one dynamic segment relaxes, and (2) the relaxed segment is paired with its molecule chain. Based on the pairing issue of transition probability, the density of probability ($p_n$) can be expressed as follows,

[Equation (11)]

where $P(A_i) = \dfrac{1}{n}$ is the transition probability for the $i$th segment in its molecule chain starting to relax, $P(A_iA_j) = \dfrac{1}{n(n-1)}$ ($i \neq j$) is the transition probability for the $i$th and $j$th two segments in their molecule chains starting to relax, $P(A_iA_jA_k) = \dfrac{1}{n(n-1)(n-2)}$ ($i \neq j \neq k$) is the transition probability for the $i$th, $j$th and $k$th segments in their molecule chains starting to relax, and $P(A_1A_2\cdots A_n) = \dfrac{1}{n!}$.

Thus, the transition probability ($P(n)$) can be obtained as,

[Equation (12)]

Based on equation (12), it is found that transition probability ($p_n$) for the case that at least one basic segment starts to relax in pairs with its corresponding molecule chain is $p_n|_{T \geq T_g} = 1 - \dfrac{1}{e}$. Then, the transition probability is $p_n|_{T_0 \leq T < T_g} = 1 - p_n|_{T \geq T_g} = \dfrac{1}{e}$ in the temperature range of $T_0 \leq T \leq T_g$, resulting into the configurational entropy ($S_c(T)$) to be $S_c(T)|_{T_g} = s^*/e$ and $\left.\dfrac{\partial S_c(T)}{\partial z}\right|_{T_g} = 0$ [35,36]. Furthermore, the domain size range is $[1, \infty)$, resulting into the value of domain size being shifted to $z(T_g)=e+1$, as shown in Figure 6. As is well known, the domain size ($z$) is determined by the configurational entropy ($S_c(T)$) and configurational number ($\Omega$), which is equal to the domain size ($z$) at $T_g$ [33,34], i.e., $\Omega(T_g)=z(T_g)=e+1\approx3.718$. The configurational entropy per mole of monomer is then obtained as $s^*(T)=k_B \cdot \ln\Omega = k_B \cdot \ln(e+1)$, resulting into the configurational entropy per mole of domain $S_c(T_g)=\dfrac{s^*(T_g) \cdot N_a}{z(T_g)}=\dfrac{\ln(e+1) \cdot k_B \cdot N_a}{e+1}=\dfrac{1.313 \cdot R}{e+1}\approx(2.936)$ J/kmol, which is in good agreement with the Wunderlich's and Chang's "universal" value of $S_c(T_g)$ =(2.9) J/kmol [37,38]. Based on the transition probability theory, the $T_g$ is the mathematical expectation, which is a numerical characteristics of the probability distribution of temperature ranged from $T_0$ to $T^*$.

[Figure 6]

According to equation (5b), it is obtained that a dynamic segment contains $N|_{T_g}$ $\approx$21 monomers into it at $T_g$, if $\chi$ =0.5. And there are $z(T_g)=e+1\approx3.718$ segments in a domain to relax due to the dynamic cooperativity and molecule entanglement. Therefore, there are $N|_{T_g} \cdot z(T_g) \approx 78$ monomers to relax synchronously at $T_g$. The claim of "50~100 monomers in a polymer chain relaxing synchronously at the glass transition temperature" is then verified using the newly proposed models.

**3. Time-space correlation between glass transition and free volume**

Based on the chaos theory, the glassy matter is classified as a dynamically chaotic system, which is always governed by deterministic laws and highly sensitive to initial conditions. The glass transition is a dynamic fluctuation of glassy matter, in which the free volume provides dimensional space for relaxation. Therefore, the connection between glass transition and free volume is governed by the constitutive relationship between dynamic chaos and mathematic fractal, i.e., the glass transition is a temporal chaos of free volume, while it is a dimensional fractal of the glass transition, as shown in Figure 7.

[Figure 7]

Studies on the glass transition of amorphous polymer, one of the most popular types of glassy matters, have been ongoing for decades since the free-volume assumption theory was formulated by Flory and Fox. This theory suggests that the thermal expansion of glassy matter involves expansions of free volume and occupied volume, which are actually occupied by the holes and molecules, respectively [39-41]. Eyring correlated the glass transition to the activation energy [42]. Doolittle and Cohen-Turnbull introduced the concept of free volume and related it to the viscosity and diffusivity parameters to predict the glass transition of glassy matter [43-45]. Fujita was the first researcher to relate free volume to molecule diffusion in glassy matter [46,47]. In the perspective, the connection between free volume and glass transition has been well-documented by Dr. White and Prof. Lipson [20].

Inspired by this work, we used a stylized depiction of the various contributions to the total volume, with a guide to notation listed below the diagram, as shown in Figure 8(a). $V$ is the total volume, the white region of $V_{\text{free}}$ is the free volume, the gray region of $V_{\text{vib}}$ is the free volume for vibrational motion and the blue region of $V_{\text{occ}}$ is the occupied volume by molecular chains. Then, the temperature-dependent volume of glassy matter is presented in Figure 8(b). At 0 K, the total volume of glassy matter is completely occupied by the molecular chains, and there is no free volume for relaxation and vibrational motion, i.e., $V=V_{\text{occ}}$. At $T_0$, the total volume of glassy matter is composed of the occupied volume ($V_{occ}$) and free volume ($V_{vib}$) for the molecular chains and their vibrational motions, respectively, i.e., $V=V_{occ}+V_{vib}$. At $T_g$, the total volume of glassy matter is composed of three parts, i.e., the occupied volume by the molecular chains ($V_{occ}$), and two types of free volumes for relaxations and vibrational motions ($V_{vib}+V_{free}$), thus, $V=V_{occ}+V_{vib}+V_{free}$, resulting in a dynamic fluctuation of total volume at $T_g$, as shown in Figure 8(c).

[Figure 8]

All these works have reported that the glass transition occurs at a specific value of free volume in the glassy matter. However, research on the connection between free volume and glass transition remains at the phenomenology and phenomenological model-building stages. Physical insights into the nature of free volume and its connection to the glass transition remain perplexed.

## 3.1 Physical insight into free volume

Dynamic glass transition can be well understood using the free-volume assumption, where the free volume is a measurement of the internal space available for the mobility freedom of dynamic segment units [20,40,41,43]. However, direct observation and examination of the free volume has not yet been achieved, and the physical insight into free volume remains at the phenomenological stage and has not been fully understood [20]. It is similar to the gaseous oxygen molecules, which do exist in water, but cannot be directly observed and touched. Therefore, the free volume has always been treated as a phenomenological and empirically determined parameter in previous models [20,40,41,43].

According to the thermodynamics of phase and microphase separations, the free volume originates from the disassociation of cohesive energy and is governed by the Einstein's mass-energy equation [32]. Below $T_g$, the relaxation motion of molecule chain results from the disassociation of intermolecular cohesion (i.e., van der Waals force) and is governed by the dynamics of phase separation. On the other hand, the relaxation motion of a segmental unit in a molecule chain is derived from the disassociation of intramolecular cohesion (i.e., covalent force) and is governed by the dynamics of microphase separation above $T_g$. At $T_g$, the glassy matter undergoes an isometric transition [40] and interplay of fluctuation in phase and microphase separations, which are determined by intermolecular and intramolecular cohesions, respectively. However, both intermolecular and intramolecular cohesions are governed by the same Einstein's mass-energy equation but with different constants. The dynamic fluctuation of glass transition in glassy matter results from the differences in disassociations of intermolecular and intramolecular cohesions at the condensed and molecular scales, respectively, resulting in a volumetric scaling effect. Therefore, there are two explicitly different constants in Einstein's mass-energy equations used to describe the intermolecular and intramolecular cohesions as a function of volume, for the glassy matter below and above $T_g$, respectively.

When segments are chemically crosslinked together to form a molecule chain, of the volume of the chain is actually smaller than the sum of the volumes of individual segments [48]. Meanwhile, when the molecule chains are packed together to form a condensed matter, the volume is actually smaller than the sum of the volumes of the individual chains [48]. The increase in volume results from the decrease in cohesive energy [48]. The corrected values for the volume and cohesive energy can be derived from the Einstein's mass-energy equation and expressed as [30],

$$[Equation\ (13a)]$$

$$[Equation\ (13b)]$$

where the constants $C_{inter}$ and $C_{intra}$ are the correction factors that Kanig introduced for the glassy matter below and above $T_g$, respectively [49]. Subscripts 1 and 2 are used to denote the initial and final states, respectively.

Furthermore, the Einstein's mass-energy equation ($E = Mc^2$, where $E$ is the cohesive energy, $M$ is the mass and $c$ is the velocity of light) is employed to construct the relationship between mass and free volume for the cohesive energy, i.e.,

$$[Equation\ (14a)]$$

$$[Equation\ (14b)]$$

Based on equation (14), there is an inverse proportion relationship between cohesive energy and volume. According to the AG model [26], the domain size ($z$) has been employed to displace the mass, i.e., $z \propto M$, resulting in,

$$[Equation\ (15a)]$$

$$[Equation\ (15b)]$$

Therefore, it is found that the change in volume is determined by the cohesive energy and governed by the Einstein's mass-energy equation. The cohesive energies result from intermolecular cohesive energy of the condensed chains and the intramolecular cohesive energy of the segmental unit in a single chain. These energies result from explicitly different constants in equation (15) below and above $T_g$, respectively.

## 3.2 Fractal insight into free volume and its connection to scaling rule

Based on chaos and fractal theory, the glass transition is a dynamic fluctuation of the glassy matter. In this fluctuation, the free volume provides the dimensional space for dynamic chaos. Therefore, the glass transition can be described as a temporal chaos of free volume, while the free volume itself is a dimensional fractal of the glass transition. The connection between free volume and the glass transition is governed by the fractal theory. This theory relies on self-consistency and power law, which are expected to support the de Gennes's scaling rule for glassy matter.

Fractals are infinitely complex, four-dimensional patterns that exhibit self-similar dynamics across different scales. They play a key role in achieving self-consistent dynamics in glassy matter. Fractals are formulated using a simple equation that is iteratively applied thousands of times, with each iteration feeding the answer back into the start of the equation as,

[Equation (16)]

where $Z$ is a complex function and $C$ is a complex constant, they both have a real part and an imaginary part. And $i$ is the exponential constant. We start by plugging a value for the variable '$C$' into the equation (16). Each complex number is actually a point in a 2-dimensional plane. The equation gives a calculation answer, '$Z_{n+1}$'. We plug this $Z_{n+1}$ back into the equation to displace of '$Z_n$', and calculate it again.

This is the fate of most starting values of '$C$'. However, some values of '$C$' do not get bigger, but instead get smaller, or alternate between a set of fixed values. These are the points inside the Mandelbrot Set [50,51]. Outside the Set, all the values of '$C$' cause the equation to go to infinity. All starting values of $C$ outside the Mandelbrot Set cause $Z$ to go to infinity. All starting values of $C$ in the Mandelbrot Set cause $Z$ to stay finite. In accordance with the fractal theory, the dynamics in glassy matter adhere to the principles of power law and self-consistency. The dynamic fluctuation of glass transition results in a scaling effect, which is also derived from the principles of power law and self-consistency. Therefore, it is essential to explore the constitutive relationship between the scaling effect and fractal theory. Below $T_g$, the dynamic glassy matter is in the form of a cubic system, with a fractal dimension above 2 due to its symmetry [52]. Above $T_g$, the dynamic glassy matter can also be modeled as a Koch fractal, with a fractal dimension smaller than 2. Therefore, the difference in fractal dimension determines the dynamic fluctuation and scaling effect in glassy matter, while the underlying dynamics are still governed by the principles of power law and self-consistency in the fractal theory [52].

### 3.3 Dynamic equilibria of free volume at glass transition

For the glassy state of amorphous polymer, an isometric free volume is expected to occur during the dynamic glass transition at $T_g$ [40]. Therefore, the critical value of free volume is expected to facilitate the glass transition [40]. The Flory-Huggins (FH) solution theory has been introduced to describe the free volume, in which the holes act as solvent molecules in glassy systems. Here the thermodynamics of glassy matter, which comprises amorphous polymer and free volume, is governed by the FH solution theory, which formulates the dynamic equilibrium of free volume. Based on the FH solution theory, the free volume acts as the poor solvent, and its amount is then decreased, resulting in the phase separation of glassy matter during its cooling process from above to below $T_g$. On the other hand, the free volume acts as the good solvent, and its amount is then increased, resulting in the microphase separation of glassy matter during the heating process from below to above $T_g$.

We assume glassy matter to be a mixture of two interconvertible phases, e.g., a dilute phase of glassy matter and free volume, as well as a dense phase of glassy matter. The fractions of dilute and dense phases are denoted by $\varphi_1$ (or $\varphi$) and $\varphi_2$ (1-$\varphi$), respectively, as shown in Figure 9(a). According to the dynamic equilibrium of phase separation, the Gibbs free energy per molecule ($G$) of an athermal non-ideal mixture is the sum of the contributions from both phases,

$$[\text{Equation (17)}]$$

where $G_1$ and $G_2$ are the Gibbs free energies per molecule of dilute and dense phases, respectively, $k_B$ is the Boltzmann constant, $\Delta S_M$ is the mixing entropy, $T$ is the temperature, and $\chi$ is the interaction parameter between two phases.

According to equation (17), the Gibbs free energy per molecule ($G$) of an athermal non-ideal mixture can be rewritten as:

$$[\text{Equation (18)}]$$

Based on equation (18), the chemical potential of two phases can be obtained,

$$[\text{Equation (19)}]$$

where $\mu_1$ and $\mu_2$ are the chemical potentials per molecule of dilute and dense phases, respectively.

Due to the dynamic equilibrium of the two phases, the chemical potential of dilute phase is equal to that of the dense one, i.e., $\mu_1 = \mu_2$, resulting in,

[Equation (20)]

Substituting equation (20) into (18) yields:

[Equation (21)]

Owing to the function of $\chi = A + B\varphi + \dfrac{C}{T}$, the equation can finally be expressed as:

[Equation (22)]

where $\Delta G = G_1 - G_2$ is the enthalpy change per molecule transformed from the dense phase to its dilute one. Meanwhile, $\chi$ is always used to present the enthalpy change based on the thermodynamics in solution theory. Therefore, the parameter of $\varphi$ is derived from the enthalpy change of glassy matter, based on the dynamic equilibrium of dilute and dense phases.

On the other hand, the specific volume ($V_1$) of dilute phase is incorporated from the mixture of glassy matter and free volume, where the fractions are $(\varphi - \varphi_f)$ and $\varphi_f$, respectively, as shown in Figure 9(b).

[Figure 9]

During the mixing of glassy matter and free volume in the dilute phase, it follows the rule of mass conservation, i.e.,

[Equation (23)]

where $\rho_1$ and $\rho_2$ are the densities of dilute phase (i.e., mixture of glassy matter and free volume) and dense phase (i.e., pure glassy matter), respectively.

Therefore, equation (23) can be rewritten as:

$$[\text{Equation (24)}]$$

In the dilute phase, there is a dynamic equilibrium of mass conservation before and after mixing between glassy matter and free volume, of which the density is assumed to be zero in comparison with that of the glassy matter, i.e., $\rho_1 V_1 = \rho_2 V_2$, thus,

$$[\text{Equation (25)}]$$

where $V_f = V_1 - V_2$ is the free volume. The $\varphi_f$ is determined by the fraction of dilute phase ($\varphi$).

Therefore, both the free volume ($V_f$) and free-volume fraction ($\varphi_f$) can be obtained based on the equations (22) and (25), in which the $\varphi$ and $\varphi_f$ are derived from the enthalpy and entropy functions, respectively.

One group of experimental data [20] of an amorphous polymer was used to verify the analytical results generated using the proposed models, based on equations (22) and (25). The parameters used in the calculations using equations (22) and (25) are listed in Table 1. Based on the experimental results [20], the density ratio of the dilute phase to the dense phase is obtained as:

$$[\text{Equation (26)}]$$

Based on the equation (24), the $\varphi_f / \varphi$ can be obtained as,

$$[\text{Equation (27)}]$$

Figure 10(a) plots the analytical and experimental results [20] of specific volume as a function of temperature, where $T_g = 374$ K. Furthermore, the constitutive relationships of dilute phase volume and free-volume concentrations ($\varphi$ and $\varphi_f$) as a function of temperature have been plotted in Figure 10(b). It is revealed that the dilute phase and free-volume concentrations ($\varphi$ and $\varphi_f$) are 7.34% and 2.48%, respectively, at the $T_g$ =374 K.

[Table 1]

[Figure 10]

## 4. Condensed constant and its connection to bivariate transition probability

Analogous to the Boltzmann constant ($k_B$ =1.3806×10$^{-23}$ J/K) and gas constant ($R$ =8.314 J/mol·K), it is expected that there is a condensed constant ($C$) to represent a constitutive relationship between temperature and activation energy for the glassy matter at the condensed-matter scale, in temperature range from $T_0$ to $T_g$. According to the previous study [19], the activation energy and volume functions play a synergetic role in determining the transition probability of glassy matter. Since the free volume concentration ($V_f$) is a given constant at $T_g$, the effect of the transition probability ($P(V)$) of volume function on the glassy matter is then kept constant. According to the hard-sphere free volume theory [44], the average distribution of the volume function ($p(V)$) in a glassy matter is associated with the free-volume concentration ($V_f$),

[Equation (26)]

where $V$ is the total volume of glassy matter and $\gamma$ ($0.5 \leq \gamma \leq 1$) is a numerical factor to correct the overlap of free volume [44].

Here the conditional probability of the volume function ($P(V)$) at $T_g$ (or $V\big|_{T_g} = V_0 + V_f\big|_{T_g}$) is then obtained [44],

[Equation (27)]

where $V_0$ is the initial total volume occupied by all the molecules, and it is kept constant as $V_f\big|_{T_g} = V\big|_{T_g} - V_0$.

Therefore, the transition probability of the volume function is obtained as $P(V)\big|_{T_g} = e^{-\gamma}$. Furthermore, the transition probability of glassy matter is a bivariate transition probability function, i.e., ($P(E_a, V)$), which involves two variables: activation energy ($E_a$) and volume ($V$) functions. Although the transition probability of activation energy ($E_a$) is governed by the Binomial distribution (or Bernoulli distribution), the sigmoid function is often used to describe the effect of activation energy ($E_a$) on the transition probability. Therefore, the conditional probability of activation energy ($P(E_a)$) in glassy matter is described by the Eyring equation [26]:

[Equation (28)]

where $\Delta E_a = E_a - E_{a0}$ is the change in activation energy at $T_g$.

Combining equations (27) and (28), we obtain the bivariate transition probability ($P(E_a, V_f)$), in which the individual probability functions of independent random variables can be expressed without reference to the other variables at $T_g$, as follows:

[Equation (29)]

For the most cases of $\Delta E_a = 0$ [26] and $\gamma = 1$ [44], equation (29) can be expressed as,

[Equation (30)]

Actually, equation (30) is able to support the equation (12), that the transition probability is $P(E_a, V)\big|_{T=T_g} = P(n)\big|_{T_0 \leq T < T_g} = 1 - P(n)\big|_{T \geq T_g} = \dfrac{1}{e}$.

In another case of $\Delta E_a > 0$ and $\gamma \in \left[\dfrac{1}{2}, 1\right)$, equation (30) can be expressed as

follows:

[Equation (31)]

Here, the condensed constant ($C$) is introduced, where $CT$ ($T_0 \leq T \leq T_g$) is used to represent the activation energy of glassy matter at the condensed-matter scale, i.e., $E_a(T) = C \cdot T$ ($T_0 \leq T \leq T_g$). Thus, equation (31) can be rewritten as:

[Equation (32)]

where $E_a|_{T_g} = CT_g$ and $C_{inter} = 1.73$ [32].

Therefore, the condensed constant ($C$) is expressed as:

[Equation (33)]

where $\varphi_f = 2.48\%$ and $R = 8.314$ J/(mol·K).

Finally, the condensed constant is obtained as $C = 0.12 \cdot (1-\gamma)$ (J/(mol·K)). As in Figure 11, the condensed constant is a given constant to represent that the activation energy of condensed matter is only determined by the temperature, i.e., $E_a(T) = C \cdot T$ ($T_0 \leq T \leq T_g$), analogous to that of the Boltzmann ($k_B$) and gas ($R$) constants for molecular dynamics and continuum mechanics, respectively.

[Figure 11]

## 5. Conclusions

This study aims to answer the question of "what is the nature of glassy state?", which is one of the greatest scientific conundrums in condensed-matter physics. The reconciliation of physics and chemistry, which meet at the dynamic glass transition of glassy matter had been achieved to unify the laws of physics at condensed-matter and molecular scales, resulting in the dynamic fluctuation and scaling effect. Three main contributions can be summarized as follows: (1) Firstly, the glassy matter is a dynamically condensed matter where physics meets chemistry at the glass transition. The glass transition is a dynamic fluctuation of glassy matter, which results in a scaling effect due to the interplay of interdisciplinary complexity at condensed and molecular scales. There are two dynamic equilibria for the glass transition, i.e., the dynamic equilibrium of order-to-disorder free-energy, and the dynamic cooperativity of domain size for molecule entanglement. It has been verified that nearly 78 monomers relax cooperatively and synchronously at $T_g$. (2) Secondly, the connection between the glass transition and free volume is derived from dynamic chaos and mathematic fractal. The scaling effect in the volume function results from the dynamic fluctuation in chaos and dimensional difference in fractals. The physical insight into free volume is governed by Einstein's mass-energy equilibrium. Fractal insight into the free volume is governed by self-consistent iteration, resulting in a time-space correlation between the glass transition and free volume. Phenomenological free-volume fraction of 2.48% is kept constant at $T_g$, according to the dynamic equilibria of enthalpy and entropy in FH solution theory. (3) Thirdly, the condensed constant ($C$) has been obtained using the bivariate transition probability function of activation energy and volume functions, where the transition probability of the volume function is kept at a constant value of $e^{-\gamma}$, and the transition probability of activation energy is also determined by the free-volume concentration according to the Einstein's mass-energy equilibrium. Based on the bivariate transition probability ($e^{-1}$) of activation energy and volume functions at $T_g$, the condensed constant ($C$) is obtained as $0.12 \times (1-\gamma)$ J/(mol·K), analogous to that of the Boltzmann ($k_B$) and gas ($R$) constants. It is expected that the activation energy ($E_a$) of glassy matter is only proportional to the temperature ($T$) at the condensed-matter scale, i.e., $E_a = CT$ ($T_0 \leq T \leq T_g$), where the condensed constant ($C$) is kept constant.

We have further identified the question of "Can the laws of physics be unified?", which has been addressed by the reconciliation of physics and chemistry. The scaling effect is governed by the interplay of interdisciplinary complexity at the condensed-matter and molecular scales, which are ruled by the thermodynamics of inter- (phase separation) and intra-molecular (microphase separation) cohesions, respectively. These effects result from the dynamic fluctuation of the glassy transition. Moreover, based on fractal theory, the Pierre-Gilles de Gennes's scaling law is derived from the dimensional differences. The mathematical iteration follows the power law and self-consistency.

**(iii) information about yourself**

Dr. Haibao Lu received his Doctor of Philosophic degree and full professor position in mechanics at Harbin Institute of Technology in 2010 and 2012, respectively. Since 2007, he has authored and coauthored more than 180 papers in refereed international journals, as well as edited seven books and book chapters. Dr. Lu has conducted foundational research on polymer physics and soft matter, including glass transition, scaling effect, mechanochemical complexity, and mechanics of soft matters.